\documentclass[article]{JHEP3}
\usepackage{cite}
\usepackage{epsfig}
\usepackage{amsfonts}

\newcommand{\be}{\begin{equation}}
\newcommand{\ee}{\end{equation}}
\newcommand{\bea}{\begin{eqnarray}}
\newcommand{\eea}{\end{eqnarray}}

\newcommand{\bastar}{\begin{eqnarray*}}
\newcommand{\eastar}{\end{eqnarray*}}

 \def \H {\mathbb{H}}

\title{Accelerating cosmologies from exponential potentials}

\author{
Ishwaree P Neupane \\ The Abdus Salam ICTP, Strada Costiera,
11-3400, Trieste, Italy\\Department of Physics,
    National Taiwan University, 106 Taipei, Taiwan, and\\
    Central Department of Physics, Tribhuvan University,
    Kathmandu, Nepal\\
E-mail: \email{ishwaree.neupane@cern.ch,
ishwaree@phys.ntu.edu.tw}}

\abstract{ An exponential potential of the form
    $V\sim \exp(-2c\,\varphi/M_p)$ arising from the hyperbolic or
    flux compactification of higher-dimensional theories is of interest
    for getting short periods of accelerated cosmological expansions.
    Using a similar potential but derived for the combined
    case of hyperbolic-flux compactification, we study the
    four-dimensional flat (and open)
    FLRW cosmologies and give analytic (and numerical) solutions with
    exponential behavior of scale factors. We show that, for the
    M-theory motivated potentials, the cosmic acceleration of the
    universe can be eternal if the spatial curvature of the
    4d spacetime is negative, while the acceleration is only transient
    for a spatially flat universe. We also briefly comment on
    the size of the internal space and its associated
    geometric bounds on massive Kaluza-Klein excitations. }

\preprint{\hepth{0311071} \\ IC/2003/148}

\begin{document}

\section{Introduction}
\label{sec-intro}

    A tentative acceptance of cosmic acceleration of the
    universe~\cite{astro1} has stimulated many works on string
    or M theory cosmology~\cite{Cornalba02a,Townsend03a,CHNW03b,
    Ohta1,Emparan03a,MNR1,Ohta03b,GKL03a,CHNOW03b,MNR2,Townsend03b,IPN03c,Misc2}.
    Though it is not explored yet whether string or M-theory
    predicts the distinctive features of our universe, like
    a spatially flat (or open) four dimensional expanding universe
    with small matter fluctuations, most physicists would believe
    that a four dimensional cosmology should be derived from a
    fundamental theory of gravity and particle interactions. This
    fundamental theory could be a theory of
    supergravity or superstring origin, or even a theory without
    the need of extra dimensions or supersymmetry (e.g. loop quantum
    gravity). Nonetheless, it is quite interesting that cosmic
    acceleration can arise from compactifications of higher-dimensional
    theories on hyperbolic spaces without or with
    fluxes~\cite{Townsend03a,CHNW03b,Ohta1,Emparan03a,CHNOW03b}.
    These solutions, along with their many generalizations
    ~\cite{MNR1,Ohta03b,GKL03a,CHNOW03b,MNR2,Townsend03b,IPN03c}
    deal with time-dependent scale factors of internal spaces, thereby
    overcoming a difficulty of ``no-go theorem'' given for a de Sitter
    type compactification in supergravity theories with static (and
    warped) extra dimensions~\cite{nogo1}.

    When studying the dynamics of the inflationary universe, in
    a four-dimensional effective theory, one considers
    the Lagrangian density
    \begin{equation}\label{action1}
      {\cal L} = \sqrt{-g}\, \left(
      \frac{M_P^2}{2}\,{\cal R}[g]-(\partial\varphi)^2-2V(\varphi)\right)\,
    \end{equation}
    with canonically normalized kinetic and potential terms.
    Of particular importance is the choice of the
    potential $V(\varphi)$ whose form depends on particle physics
    or compactifications used to derive a lower dimensional effective
    field theory. One would like to derive
    potentials that allow inflation in the early universe and/or acceleration in
    the current cosmological epoch. Both cases may be explained in
    a typical manner by postulating some suitable scalar potentials.

    Recently some progress has been made about inflation within
    string theory~\cite{KKLT} by keeping the system in the string
    theory de Sitter vacuum~\footnote{However, this minimum
    is only metastable. There are arguments~\cite{Giddings03a}
    that the types of potentials one may get in warped de Sitter
    string theory geometries, with trapped fluxes present in the
    extra dimensions, can lead to instability of our four-dimensional
    world.} and by constructing a scalar potential in the
    $D3/\bar{D}3$ warped brane background~\cite{Kachruetal}.
    These mathematical frameworks assume that the metric of a
    compact non-singular internal $m$-manifold $\Sigma_{m}$ is
    time-independent, and in turn, inflation is non-generic which
    involves some degree of functional fine tuning. In this paper
    we consider scalar fields in cosmology by allowing the metric
    on $\Sigma_m$ to be time-dependent.

    For many classical compactifications of effective supergravity
    theories, the scalar potentials are of the form $V\sim
    \exp (-2c\,\varphi/M_P)$ with the (dilaton) coupling constant
    $c\sim {\cal O}(1)$. The cosmological models with such a potential
    have been known lead to interesting physics in a variety of context,
    ranging from existence of accelerated expansions~\cite{Halliwell87a} to
    cosmological scaling solutions~\cite{Ratra88a}, see
    also, e.g.,~\cite{scaling,IMS03a} for some related discussions.
    It is also worth noting that an attractor solution with a simple
    exponential potential of the above form may lead to cosmic
    acceleration for natural values of model parameters~\cite{datafit}.

    A number of authors have studied the effects of exponential
    potentials on M-theory cosmology, including the recent papers
    concerning hyperbolic or flux compactification and accelerated
    expansion. So far no general discussion has been given for the
    combined effects of hyperbolic-flux compactification, see, however,
    Refs.~\cite{Emparan03a,MNR2} for some qualitative arguments.
    In addition, most authors have studied only a spatially flat
    universe, and this in turn leads to the observation that the
    accelerated period in these models is typically preceded by
    a period when the universe was dominated by the kinetic energy
    of the scalar field. So in this note we allow the spatial
    curvature ($k$) of the 4d spacetime to take the value
    $k=0$ or $k=\pm 1$. We show that the accelerated expansion of our
    universe may continue forever if $k=-1$. This possibility was
    pointed out in~\cite{CHNOW03b,Halliwell87a}; here we make the
    earlier discussions more precise with further analysis and new
    results, taking into account the combined effects of hyperbolic
    extra dimensions and background fluxes.

\section{Compactification with fluxes}

    Let us consider that the M theory spacetime is the product of a
    four dimensional spacetime ${M}_{4}$ and an internal compact space
    $\Sigma_{m, k_1}$, namely,
    \begin{equation}\label{metric1}
    ds_{4+m}^2 = {\rm e}^{-m\phi(\xi)}\, g_{\mu\nu}(x) dx^\mu dx^\nu
    + r_c^2\,{\rm e}^{2\phi(\xi)}\,d\Sigma^2_{m, k_1}
    \end{equation}
    with $\tilde{R}_{ab}(\Sigma_{m, k_1})=k_1(m-1)
    \tilde{g}_{ab}\,r_c^{-2}$, where $r_c$ is the radius of curvature
    of internal space ($a, b$ go over the dimensions of the
    internal space). In suitable coordinates,
    $k_1=-1,\,0,$ or $ +1$, respectively, for
    hyperbolic, flat or spherical manifold~\footnote{By saying a
    hyperbolic manifold one would mean the space admitting constant
    negative curvature. A compact hyperbolic manifold (CHM) is
    achieved by taking a quotient $H^m/{\Gamma|_{free}}$ of the
    non-compact hyperbolic space $H^m$ by a freely acting discrete
    subgroup $\Gamma$ of the isometry group.}.
    Since the metric~(\ref{metric1}) is written in Einstein conformal
    frame, the effective 4d Planck mass $M_P$ does not depend on time scale
    but (only) on a spatial volume of the highly curved internal manifold,
    the latter scales as $\sim r_c^m\,e^{\alpha}$, where $\alpha$ depends
    on the topology of the internal manifold~\cite{Starkman00a}.

    Together with the four-form field strength
    $_*F_{[4]}=2b\, \mbox{Vol}(\Sigma_{m, k_1})$, $b$ being the field
    strength parameter, upon the dimensional reduction we
    get~\cite{Emparan03a,IPN03c}
    \begin{equation}\label{action2}
    I= {M_P^2} \int d^4x \left(\frac{{\cal R}[g]}{2}- \frac{m(m+2)}{4}
    \left(\partial\phi\right)^2 - b^2 {\rm e}^{-3m\phi} +
    k_1\,\frac{m(m-1)}{2\,r_c^2}\, {\rm e}^{-\,(m+2)\phi}\right),
    \end{equation}
    where $M_P^{-1}=\sqrt{8\pi G}\equiv \kappa $ is the reduced Planck
    mass. The solution following from $(4+m)$ dimensional field
    equations is
    \begin{equation}
    ds_{4+m}^2 = {\rm e}^{-m\phi(\xi)}
    \left(-\,S^{6}\,d\xi^2+S^2\,d\vec{x}_{3,k}^2\right) + r_c^2\,{\rm
    e}^{2\phi(\xi)}\,d\Sigma^2_{m, k_1},
    \end{equation}
    \begin{equation}
    \phi(\xi)= \frac{\ln(A\,B)}{m-1}, \quad S^2=
    {[A]}^{{m}/{m-1}}\,{[B]}^{(m+2)/3(m-1)},
    \end{equation}
    \begin{eqnarray}
    &&  A(\xi) = \left\{\begin{array}{l}
    \frac{r_c}{(m-1)}\,\frac{\lambda\,\beta}
    {\sinh\left[\lambda\beta\, |\xi|\,\right]}, \quad ~~ k_1=-1,\\
    {\rm e}^{\lambda\beta\,\xi},  \quad ~~~~~~~~~~~~~~~ k_1=0,\\
    \frac{r_c}{(m-1)}\,\frac{\lambda\,\beta}
    {\cosh\left[\lambda\beta\, \xi\,\right]}, \quad ~~~ k_1=+1,
    \end{array} \right.\  \nonumber \\
    &&    B(\xi) = \left\{\begin{array}{l}
    {\rm e}^{-3\lambda\,\xi}, ~~~~~~~~~~~~~~~~~ \quad b=0 , \\
    2b\,\sqrt{\frac{m-1}{2m}}\,\frac{\cosh
    3\lambda\,\xi}{\lambda\,\beta},~~~
    \quad b \neq   0 ,
    \end{array} \right.\
    \end{eqnarray}
    up to a shift of $\xi$ around $\xi=0$, and $\beta\equiv
    \sqrt{\frac{3(m+2)}{m}}$. The constant $\lambda$ has dimension
    inverse of the coordinate time $\xi$. The above solutions are
    still somewhat oversimplified because of the choice $k=0$;
    however, we will accommodate the $k =-1$ case below.

Regardless of the flux value (i.e., $b>0$ or $b=0$), the $r_c>0$
solution with $k_1=\pm 1$ does not simply reduce to the one with
$k_1=0$, where the internal space is flat (see, also the
discussions in~\cite{MNR1,CMChen02a}). For a small background
flux, namely, $b<< r_c^{-1}$, it is possible to suppress the
growth in the size of internal space, although the parameter $b$
has no much effect to the acceleration of a flat FLRW
universe~\cite{Ohta1,MNR1}.

\subsection{Field equations and 4d cosmology}

In the above section, we wrote the solutions in terms of $(4+m)$
dimensional coordinate time $\xi$. Of course, one can define a 4d
proper time $t$ using $dt = S^3\,d\xi $, but from a viewpoint of
4d cosmology it would be more desirable to express the solution
directly in terms of the 4d proper time $t$. Hereafter, we thus
take the four-dimensional part of $(4+m)$ Einstein-frame metric to
be the usual FLRW spacetime in the standard coordinate (see, 
e.g.,~\cite{CHNW03b}):
    \begin{equation}
      ds_{4+m}^2={\rm e}^{-m\phi}\left(
      -\,dt^2+a(t)^2\left(\frac{dr^2}{1-k r^2}+ r^2
      d\Omega_{2}^2\right)\right)+ r_c^2\,{\rm e}^{2\phi} d\Sigma_{m,k_1}^2,
    \end{equation}
  where $\phi=\phi(t)$. Here $k=0, \pm 1$ is the spatial curvature of the
  universe, $d\Omega^2$ the metric on a unit
  $2$-sphere, and $a(t)$ is the scale factor of the physical $3$-space.

 Next, we define a canonically normalized
    4d scalar $\varphi~(\equiv \varphi(t))$ via
\begin{equation}
    \frac{\varphi}{M_P}=\sqrt{\frac{m(m+2)}{4}}\,{\phi}
     -\frac{1}{2}\,\sqrt{\frac{m}{m+2}}\,\ln \frac{m(m-1)}{4},
     \end{equation}
so that the effective 4d Lagrangian takes the form
of~(\ref{action1}) with potential
    \begin{equation}\label{Vtotal}
     V(\varphi) = -\,k_1\,{M_P^2}{r_c^{-2}}\,
     {\rm e}^{- 2\,c\,\kappa\,\varphi}+
     \frac{1}{2}\,{M_P^2 \tilde{b}^2}\,
     {\rm e}^{-(6/c)\,\kappa\,\varphi},
    \end{equation}
    where $\tilde{b}^2\equiv b^2
    \left({4}/{m(m-1)}\right)^{3/c^2}$.
    As is evident from the result in section 2,
    the ratio between $1/r_c$ and $\tilde{b}$ may have
    an important impact on the scale factor of internal
    manifold, as well as on the potential. In order to allow
    an accelerated expansion, $V(\varphi)$ needs to be positive.
    This requires, for $k_1=+1$, $\tilde{b} r_c>\sqrt{2}$
    and $c\geq \sqrt{3}$. However, in time-dependent backgrounds,
    it turns out that the most interesting case is
    $k_1=-1$, so called {\it hyperbolic compactification}, where
    $V(\varphi)>0$ even if $\tilde{b}=0$. And, $k_1=0$ corresponds 
    to a pure flux compactification, where the internal space is 
    Ricci flat.

    To explore the inflationary universe with a scalar field,
    it may sometimes be useful not to restrict the coupling
    $c$ (other than that $c>0$), because we do not know yet
    which compactification of string/M theory, if any, best
    describes the very early universe or the observed cosmic
    acceleration. For many
    (classical) compactifications of supergravity theories,
    only $c\gtrsim 1$ arises in
    practice. For the hyperbolic compactification,
    since $c=\sqrt{\frac{m+2}{m}}$, one has $1 \lesssim c<\sqrt{3}$
    when $m\geq 2$. ($c$ here is $1/\sqrt{2}$ times the
    $c$ defined in~\cite{CHNOW03b}, but is the same
    of~\cite{IPN03c}.)

    The wave equation for $\varphi$ and the Friedman equation
    have the following form:
    \begin{eqnarray}\label{waveeqn}
     \ddot{\varphi}+3H\dot{\varphi} +
     \frac{2 k_1}{r_c^2}\, c {\rm e}^{-2c\,\varphi}
      -\frac{3}{c}\,\tilde{b}^2\, {\rm e}^{-(6/c)\varphi}&=&0\,,\\
      3H^2-\dot{\varphi}^2 +\frac{3k}{a^2}
      +\frac{2k_1}{r_c^{2}}\, {\rm
    e}^{-2c\,\varphi}-\tilde{b}^2 {\rm
    e}^{-(6/c)\varphi} &=& 0,\label{Friedman}
    \end{eqnarray}
    (in units $\kappa=1$).
    As usual $H=\dot{a}/a$ is the Hubble
    parameter and an overdot denotes differentiation with respect to
    the proper time $t$. There is an equation (of higher dimensional
    origin) which does
    not depend on the background
    fields and the (dilaton) coupling constant $c$, namely,
    \begin{equation}\label{identity}
      H^2-\frac{\ddot{a}}{a}+\frac{k}{a^2}={\dot\varphi}^2 \,.
    \end{equation}
    In our analysis, this equation is useful for consistency checks. The
    solution of equations (\ref{waveeqn})-(\ref{identity}) following from
    the 4d Lagrangian~(\ref{action1}), with the
    potential~(\ref{Vtotal}), is the same as the one from
    $(4+m)$ dimensional field equations.

    \subsection{Hyperbolic flux compactification}

     In general, for the
    acceleration of the scale factor $a(t)$, that corresponds to our
    universe, the spatial curvature $k$ has to be zero (or
    negative)~\footnote{
    See also a related discussion in Ref.~\cite{Burgess03a}, which
    discusses some cosmological implications of S-branes.}.
    Let us first consider that the four
    dimensional universe is spatially flat (i.e. $k=0$). In this case
    it is convenient to introduce a new logarithmic time
    variable $\tau$, which is given by
    \begin{equation}
      d\tau={\rm e}^{-c\,\varphi}\, dt\,, \quad
      \alpha(\tau)=\ln(a(t))\,.
    \end{equation}
     Then, equations~(\ref{waveeqn}) and (\ref{Friedman}), with
     $k_1<0$ (for simplicity, we take $k_1=-1$), take the
     following form:
     \begin{eqnarray}
        \varphi^{\prime\prime}-c\,{\varphi^\prime}^2 +3{\alpha^\prime}
       {\varphi^\prime}-\frac{2c}{r_c^{2}}-\frac{3}{c}\,\tilde{b}^2\,
     {\rm e}^{2(c^2-3)\varphi/c} &=& 0 \,,\label{hyflux1}\\
     3{\alpha^\prime}^2-{\varphi^\prime}^2- \frac{2}{r_c^{2}}-\tilde{b}^2
       {\rm e}^{2(c^2-3)\varphi/c} &=& 0 \,.\label{hyflux2}
     \end{eqnarray}
     A prime indicates differentiation with
     respect to $\tau$. The conditions for expansion and
     accelerated expansion are, respectively,
     \begin{equation}
       H=\alpha^\prime(\tau) {\rm e}^{-\,c\,\varphi} >0\,, \quad
       \frac{\ddot{a}}{a}=e^{-2c\varphi}
       (\alpha^{\prime\prime}-c\alpha^\prime \varphi^\prime
       +{\alpha^\prime}^2)>0\,,
     \end{equation}
     and so $\alpha^\prime>0 $ corresponds
     to expanding cosmologies with $H>0$.
     When $\tilde{b}= 0$ and
     $c<\sqrt{3}$, the solution, up to a shift of $\tau$ around
     $\tau=0$, is
     \begin{eqnarray}
      \sqrt{3}\,\alpha(\tau) &=& {\delta_-}\,
       \ln\cosh \left(\frac{\gamma\tau}{r_c}\right)
       +{\delta_+}\,
       \ln\sinh \left(\frac{\gamma\tau}{r_c}\right)
       +c_1\,,\label{b=0alpha}\\
      \varphi(\tau) &=& {\delta_-}\,
       \ln\cosh \left(\frac{\gamma\tau}{r_c}\right)
       -{\delta_+}\,
       \ln\sinh \left(\frac{\gamma\tau}{r_c}\right)
       +c_2\,,\label{b=0varphi}
     \end{eqnarray}
     where $c_1$ and $c_2$ are some (integration) constants and
     $ \delta_{\pm} = 1/(\sqrt{3}\pm c)$,
       $\gamma=\sqrt{(3-c^2)/2}$.
     The solution first found in~\cite{Townsend03b} is easily recovered
     by taking $r_c=1$. However, in order to keep
     the discussion sufficiently general, we shall
     keep $r_c$ as a free parameter, which is actually related with
     the size of the compact internal manifold.

     Hereafter we often use the notation $r_c^{-1}\equiv M$.
     The acceleration parameter is
      \begin{equation}
       \frac{\ddot{a}}{a}= 2M^2{\gamma}^2
       {\rm e}^{-2c\,\varphi}
      \left[\frac{2(c^2-1)}{c^2-3}
    +\frac{2\sqrt{3}c\left(2\cosh^2 {\gamma {M} \tau}-1\right)
      -c^2-3}{3(3-c^2)\left(\cosh^2 {\gamma {M} \tau}-1\right)
      \cosh^2 {\gamma {M} \tau}}\right]\,.
      \end{equation}
     From this we clearly see that, when $c>1$, accelerated expansion of the 4d
     spacetime is only transient but it can be eternal
     for the value $c\leq 1$. Especially, for the critical
     value $c= 1$, the conditions for expansion and accelerated
     expansion are satisfied when $M\tau>0.27 $. Thus the onset time
     for cosmic acceleration of physical $3$-space could depend on
     the curvature radius of internal space, $r_c\equiv M^{-1}$.
     We discuss this point further below.

     From~(\ref{hyflux1}), (\ref{hyflux2}), it appears that the
     choice $c=\sqrt{3}$ (which may arise
     from a compactification of 5d
     gauged supergravity~\cite{Paul04a}) is special. When $c\approx \sqrt{3}$,
     the solution is
     \begin{equation}
       \alpha^\prime= \frac{\tau}{2}\left(\tilde{b}^2+2M^2\right)
    +\frac{1}{6\tau},
       \quad \varphi^\prime=\frac{\sqrt{3}\,\tau}{2}
       \left(\tilde{b}^2+2M^2\right)
     -\frac{1}{2\sqrt{3}\tau}.
     \end{equation}
     As $\tau\to 0$, $t\to 2\sqrt{\tau}$,
     $H\propto 1/(6\sqrt{\tau})$
     and $\ddot{a}/a\propto -\,1 /(18\tau)$. In this case, the
     universe is expanding
     but decelerating when $\tau\to 0_+$. After a certain proper time,
     the ratio $\ddot{a}/a$ approaches zero, thereafter the universe
     accelerates in the interval $$
     2-\sqrt{3} < 3\left(\tilde{b}^2+2M^2\right)\,\tau^2 <
     2+\sqrt{3},$$
     while $\varphi^\prime$ changes its values from $\varphi_1^\prime$
     (negative) to $\varphi_2^\prime$ (positive), where
    $$ \varphi_1^{\prime}=
     -\,0.52 \sqrt{\tilde{b}^2+2M^2}, \quad
     \varphi_2^{\prime}=+\,1.41
     \sqrt{\tilde{b}^2+2M^2}.$$
     As in the scenario of~\cite{Emparan03a}, the universe
     accelerates around the turning point where
     $\varphi^\prime$ changes its sign. In turn, the acceleration
     is transient and leads to only a few e-foldings.

    Numerical solutions may be explored to verify the existence of
    transiently accelerating expansions, especially, in the $k=0$ case,
    with $c>1$. For example, as shown in
     figures~\ref{figure1}
     and \ref{figure2},
     for the following initial conditions:
     $$
     a_0=2.6\,,\quad \varphi_0=0.4\,, \quad \dot{\varphi}_0 =0.5, $$
     the solutions are only transiently
     accelerating. The late-time behavior of solutions and the period
     of accelerations are less sensitive with the initial
     conditions~\footnote{However, the
     initial behavior of the solutions is found bit sensitive
     with the initial values $a_0$ and $\varphi_0$, other than
     that with $M$. In any case, for the $k=0$ solution,
     $a$ and $\varphi$ involve some arbitrary (integration)
     constants, like $c_1$ and $c_2$ in Eqs.~(\ref{b=0alpha}),
     (\ref{b=0varphi}), which may be suitably chosen.}.

    \FIGURE[ht]{
     \epsfig{figure=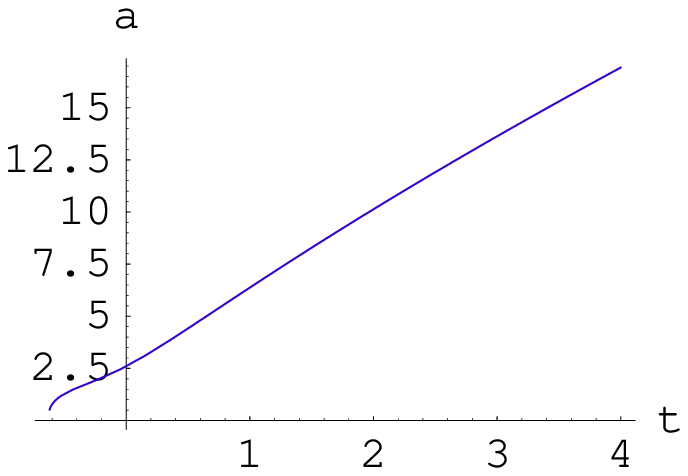, height=3.5cm, width=6.0cm}
     \hskip1.5cm
     \epsfig{figure=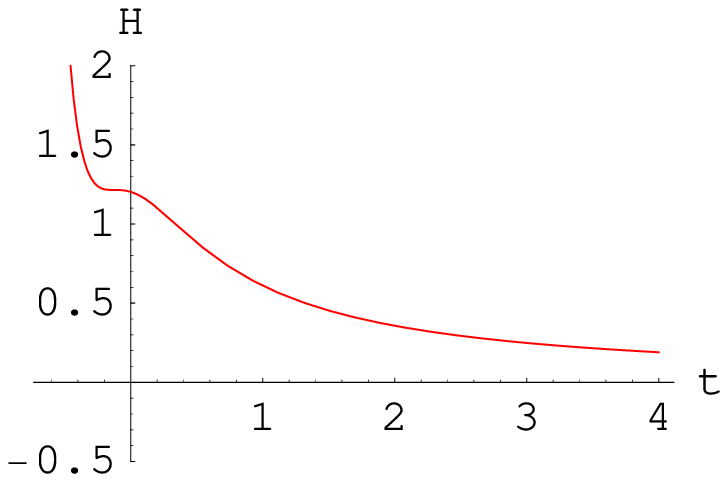, height=3.5cm, width=6.0cm}
     \caption{(Left panel) scale factor vs cosmic time $t$,
        (right panel) Hubble parameter $H$ vs $t$.
        The parameters are fixed at $c=3/\sqrt{7}$,
       $r_c=0.4$, $\tilde{b}=1$,
       $k=0$, $k_1=-1$.}
     \label{figure1}}

    From figures~\ref{figure1} and \ref{figure2}, we clearly see that the
     scale factor $a(t)$ can grow with $t$ much faster than the scale
     factor (or size) of the internal space $R_c~(\equiv r_c\times
     {f})$, where~\footnote{We have $R_c \simeq r_c$ when
     $\varphi<<M_P$.
    If $M\sim 10^{15}\,\mbox{GeV}$, then for a
    physically interesting case, the growth factor ${f}$
     should not exceed $10^{13}$ (a smaller value is more
     preferred) when the scale factor of our
    universe grew by a factor of at
    least $10^{61}$, i.e., from the Planck size $10^{-33}\,\mbox{cm}$
    to the currently observable size $\gtrsim
    10^{28}\,\mbox{cm}$.},
    \begin{equation}
    {f}={f}(\varphi(t))= \left(\frac{m(m-1)}{4}\right)^{1/(m+2)}
    \exp\left(\frac{2}{\sqrt{m(m+2)}}\,
    \frac{\varphi}{M_P}\right)
    \end{equation}

\FIGURE[ht]{
 \epsfig{figure=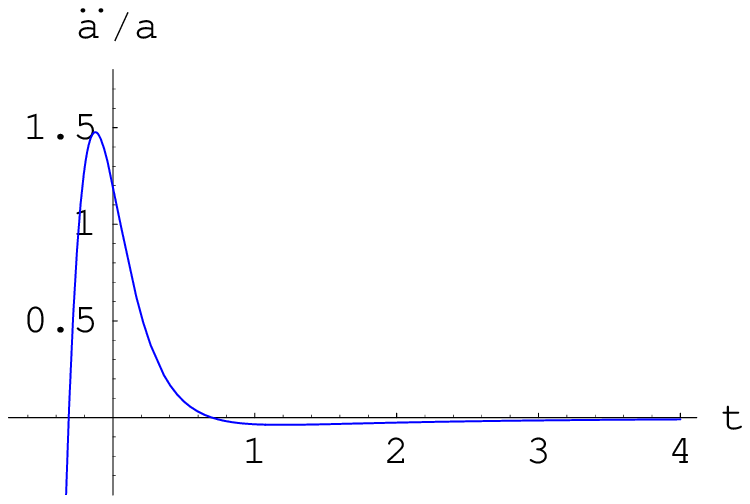, height=3.5cm, width=6.0cm}
     \hskip1.5cm
     \epsfig{figure=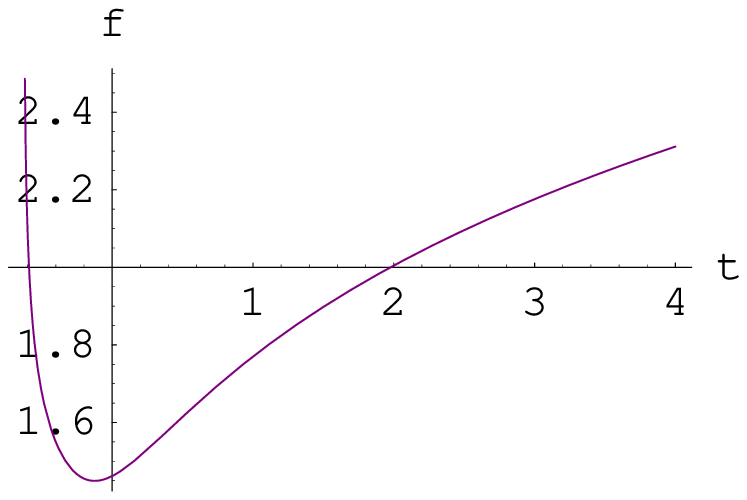, height=3.5cm, width=6.0cm}
      \caption{(Left panel) Acceleration $\ddot{a}/a$ vs
       cosmic time $t$, and (right panel) the growth factor
       $f \equiv {f}(\varphi(t))$, which measures the increase in
      the size (or scale factor) of the internal space with $t$.
      The parameter values are the same as the ones in
       Fig.~\ref{figure1}.} \label{figure2}}

     The acceleration of the 4d universe
     is associated to a bounce of the compact internal space off its
     minimal scale factor (see also a discussion in~\cite{Townsend03b}).
     Table 1 shows how the scale factor $a(t)$ and the term
     $f(\varphi(t))$ associated with the size of the internal space
     could vary with $t$, for a certain choice of initial conditions:
     $\varphi_0=0.5$, $\dot{\varphi}_0=0.5$ and 
     $a_0=2.6$~\footnote{In our analysis, we do not
     normalize the scale factor $a(t)$ to be unity, since we are
     interested to measure the rate of expansion of scale
     factors associated with the physical $3$-space and the internal manifold}.
    The other parameters are fixed at $k=0$, $k_1=-1$,
    $c=3/\sqrt{7}$ and $r_c=0.4$~\footnote{One should not be confused
    here that by taking $r_c<1$ it would mean $r_c< l_P$ (Planck length).
    $r_c$ is a free parameter whose value depends on how
    one normalizes the Ricci curvature, namely,
    ${\cal R}(\Sigma_{m,k_1})=m(m-1)k_1/r_c^2$, and a hyperbolic manifold
    simply means that $k_1<0$. So, in suitable
    coordinates, we may take, e.g., $k_1=-2$, instead of $k_1=-1$ and
    shift $r_c$ accordingly. In fact, in most of our
    discussion below we assume that $r_c\gtrsim 10^3\,l_P$,
    thus we have $r_c$ greater than the Planck length.}.
    One also finds that for $t>>1$ (i.e., $\varphi>>1$), the flux
    contribution to the acceleration or the scale
    factor $a(t)$ of physical $3$-space is almost negligible, even
    if $\tilde{b}>> r_c^{-1}$, except for the case that $c\sim \sqrt{3}$.

     \begin{table}[ht]
       \label{table1}
    {\bf Table 1: } The scale factor $a(t)$ and $f(\varphi(t))$
    as the function of proper time $t$.
\begin{center}
\begin{tabular}{|c|c|c|c|c|} \hline
$ {} $ & $t={10}^2$  &  $t={10}^4$ & $ t={10}^6$ & $t={10}^8$\\
\hline

\hline $a $ & $  2.1\times 10^{2} $ & $ 7.2\times {10}^3 $ &
$3.8\times {10}^5 $ & $8.6\times {10}^6$
\\
\hline ${f} $ & $ 4.7 $ & $ 13 $ & $35  $ & $100$
\\ \hline
\end{tabular}
\end{center}
\end{table}

     The combined effects of hyperbolic-flux
     compactification\footnote{More generally,
     M-theory compactification since the form-field
     background
     is typical in M-theory.} look different in many ways from the
     flux or hyperbolic compactification alone.
     This behavior can be seen also from the solutions
     presented in section 2. For example, if the curvature
     radius $r_c$ is sufficiently small and the flux
     parameter is $\tilde{b}\gtrsim r_c^{-1}$, then the acceleration
     parameter $\ddot{a}/a$, although negative, is almost
     zero for all $t$. That is, the deceleration following after a
     period of transient acceleration can be modest.
     In our model, for a spatially flat 4d universe ($k=0$), the
     acceleration is only a transient phenomenon for $c>1$ but
     it may be eternal for $c\leq 1$.

     Some of the above discussions will, however, change if the spatial
     curvature of physical 3-space is negative. In this case,
    the cosmic acceleration may continue forever, even if
    $c>1$, when the background is triggered by
    non-trivial fluxes, like the
    parameter $\tilde{b}$, or by a higher dimensional
    cosmological constant. This observation is not totally new,
    which was made before in the work of Halliwell~\cite{Halliwell87a}
    but rather implicitly. We will discuss this effect more explicitly,
    in the presence of background flux, in subsection 2.4.

     To deal with the late-time acceleration,
     there may be a need
     to introduce matter fields, since the (contemporary) universe
     contains about $30\%$ (baryonic plus dark) matter.
     But neither matter nor
     radiation contribution will make much difference to the amount
     of acceleration in these models, as it is clear from the relation
     $\dot{a}^2\propto a^2\,\rho_{R, M}+ \mbox{constant} $, where
     $\rho_M\propto 1/a^3$ and $\rho_R\propto 1/a^4$;
     matter and radiation both satisfy
     the strong energy condition separately. The introduction of cold
     dark matter is phenomenologically well motivated and
     may be required for the construction of the effective 4d
     model nonetheless.


\subsection{Toroidal flux compactification}

For flat models (i.e., $k_1=0$ case) with a non-zero flux, we find
it convenient to define
      \begin{equation}
    d\tau={\rm e}^{-(3/c)\varphi}\, dt\,, \quad
    \alpha(\tau)=\ln(a(t))\,.
      \end{equation}
 Furthermore, with $k=0$, the field
      equations have the following form:
      \begin{eqnarray}
    \varphi^{\prime\prime}-c\,{\varphi^\prime}^2 +3{\alpha^\prime}
           {\varphi^\prime}&=&\frac{3\tilde{b}^2}{c}\,,\\
           3{\alpha^\prime}^2-{\varphi^\prime}^2&=&\tilde{b}^2\,.
      \end{eqnarray}
 Especially, for $c \approx \sqrt{3}$, the solution is simple,
 which is given by
      \begin{equation}
    {\rm e}^{\alpha(\tau)}\propto {\rm e}^{\tilde{b}^2\tau^2/4}\,
    \tau^{1/6}\,,   \quad  {\rm e}^{-\sqrt{3}\varphi} \propto {\rm
      e}^{-3\tilde{b}^2\tau^2/4}\, \tau^{1/2}\,. \end{equation} This
      solution is accelerating in the interval
      $(\sqrt{3}-\sqrt{2})<\sqrt{6}\,\tilde{b}\,\tau <
      (\sqrt{3}+\sqrt{2})$. Clearly, the solution with $\tilde{b}>0$
      is different from that with $\tilde{b}=0$, in the latter
      case there is no acceleration. By the same token, the
      $c\neq \sqrt{3}$ solution can be different from that with
      $c=\sqrt{3}$. For example, when $c=\sqrt{4/3}$ (i.e., $m=6$),
      the parametric solution is
      \begin{equation}
    \alpha^{\prime}= \frac{\tilde{b}\,(x^2+1)}{2\sqrt{3}\,x}
    \,, \quad \varphi^{\prime}= \frac{\tilde{b}\,(x^2-1)}{2\,x}\,,
      \end{equation}
      where $x=x(\tau)$ satisfies the equation
      \begin{equation}
    \frac{c_+}{d_+}\,\tanh^{-1} \left( \frac{x}{d_+}\right) +
    \frac{c_-}{d_-}\,\tan^{-1}
    \left(\frac{x}{d_-}\right)=3 \tau \tilde{b}\,,
      \end{equation}
      with $c_{\pm}= 4\sqrt{2}\pm 3\sqrt{3}$,
      $d_{\pm}=\sqrt{3\sqrt{6}\pm 7}$.
      The limit $x\to 0_+$ ($\varphi^\prime \to -
      \infty$) corresponds to $\tau\to 0_+$, while the limit $x\to 1$
      ($\varphi^\prime \to 0$) corresponds to $\tau\sim 1/2\tilde{b}$.
      There is a transient acceleration when $x\to 1$. Similarly, for
      $c>\sqrt{3}$, there may arise a period of acceleration for
      the $k=-1$ cosmology (see also a related 
      discussion in~\cite{Cornalba02a}).

      \bigskip


    \subsection{Open universe and cosmic acceleration}

    The recent astronomical observations~\cite{astro1} appear to indicate
    that our observable universe is well described by flat Euclidean
     geometry. Does it mean that we shall study/construct only the
     spatially flat FLRW models and postpone investigation
     of open space cosmologies? The answer is perhaps not
     affirmative, since one cannot rule out the possibility of
     having a negatively curved four-dimensional universe
     in large distance scales, see,
     e.g.,~\cite{Padma03a}~\footnote{The spatial geometry can be
     nearly flat locally even if $k=-1$. 
     The effect of the non-trivial topology
    is prominent only if the spatial hypersurface is
    smaller than the observable region at present.}.

    Let us note that the most interesting case,
      $1<c<\sqrt{3}$, is an example of
      hyperbolic compactification which gives a power-law
      acceleration. To make our
      discussion more precise, we take $k=k_1=-1$ in the equations
      (\ref{waveeqn}), (\ref{Friedman}) and set $b=0$. The background
      solution is
      \begin{equation}\label{zero-flux}
    a_0=\frac{c\,t}{\sqrt{c^2-1}}, \quad \varphi_0=\frac{1}{c} \,
    \ln\left({{M} c t}\right)\,.
      \end{equation}
      Here the branch $c<1$ can be unstable due to an
      imaginary scale factor $a(t)$, so we require
      $c \gtrsim 1$~\footnote{We do not expect that
      such a solution is responsible for the primordial inflation, so
      we shall not concern ourselves with finer details, like why
      $\Omega_\varphi<1$ and whether one can solve the flatness problem
      if $c> 1$.} and hence $\Omega_\varphi\equiv
      8\pi G\rho_\varphi/(3H_0^2)=1-1/a_0^2H_0^2=1/c^2\lesssim 1$.
      The limit $c=1$
      however may be approached once $\tilde{b}>0$.
      From~(\ref{zero-flux}),
      using $d\tau=e^{-c\,\varphi}\,dt$,
      we derive
      $t \propto {\exp}({{M} c \tau})$. In the limit
      $\tau >> {M}^{-1}$, so $t>>1$ and $\varphi>>1$,
      the contribution of the background fluxes to the potential
      is almost negligible. Indeed, since $\dot{\varphi}_0=1/(ct)$, the
      internal space never accelerates, rather its expansion rate
      decreases as $t$ advances (see also figure~\ref{figure3}).

      To lowest order, where $a=a_0+\delta a_0$
      and $\varphi =\varphi_0+\delta\varphi_0$, the field
      equations take the following form:
      \begin{equation}
    \delta\ddot{\varphi}_0 + 3 H_0 \delta{\dot{\varphi}_0} +
    3 \delta{H_0} \dot{\varphi_0}
    + {4 M^2} c^2 {\rm e}^{-2\,c\,\varphi_0} \delta{\varphi_0}
    =0
    \end{equation}
      \begin{equation}
    6H_0 \delta{H_0} - 2\dot{\varphi}_0 \delta{\dot{\varphi}_0}
    + {4 M^2} c {\rm
      e}^{-2c\,\varphi_0} \delta{\varphi_0}
      + \frac{6}{a_0^2}\,\frac{\delta {a_0}}{a_0}
    =0,
      \end{equation}
      where $\delta H_0=-\,(\delta a_0/a_0)\, H_0+ \delta \dot{a}_0/a_0
      $. These equations have the solution
      \begin{equation}\label{linearsol}
    \delta a_0= \beta\,t^n\,, \quad \delta \varphi_0=\beta \gamma\,t^{n-1} \,,
      \end{equation}
      where $\beta$ is undermined but a small constant, and
      \begin{equation}
    \gamma=\frac{3(1-n)}{4}\sqrt{c^2-1}\,,\quad n^2=
    \frac{4}{c^2}-3.
      \end{equation}
      Thus, for $c<\sqrt{4/3}$, the acceleration tends to zero
      asymptotically, and there is no cosmological event
      horizon in this case. The behavior of the solutions
      (or trajectories in the phase portraits) may change
      at $c=\sqrt{4/3}$. In fact, the perturbative solution with
      $b=0$ may give an eternally accelerating expansion only
      if $m> 6$ (i.e., $c < \sqrt{4/3}$) (which is in precise agreement
      with the earlier discussions
      in~\cite{Halliwell87a,CHNOW03b}),
      while, for $c > \sqrt{4/3}$, the late-time
      acceleration can be oscillatory; this limit, however, can be pushed to
      $c>\sqrt{2}$ with $\tilde{b}>0$.

      Let us take the point of view that a non-trivial flux
      (in extra dimensions) serves as a source term to lowest
      order perturbations of the
      scale factor and hence the late-time acceleration of the universe.
      In this case, the constant $\beta$ may be
      fixed in terms of $\tilde{b}$ and $M$, namely,
      \begin{eqnarray}
    &&\beta = {\tilde{b}^2}\,
    {M^{n-3}} \left(\frac{1}{\gamma}\right)
    \frac{(3n-1)(3-n)}{2n(3n^2+2n+3)}\,
    \left(\frac{3-n}{6}\right)^{1-n/2}\,, \nonumber \\
    && \gamma =\frac{1-3n}{6(1-n)}\,\sqrt{9-n^2}; ~~\quad n=3(1-2/c^2)\,.
      \end{eqnarray}
      The most suitable range turns out to be $1<c<\sqrt{2}$, and hence
      $$ 0<\gamma<\sqrt{8/9}.$$
      For $c=3/\sqrt{7}$ (i.e., $m=7$), one has
      $\gamma=\sqrt{7/8}$, $\beta \simeq 0.71\,\tilde{b}^2
      M^{-14/3}$.
      Here only the branch $c<\sqrt{2}$ is (eternally) accelerating. However,
      the full classical non-linear equations, with $\tilde{b}>0$, may still
      allow a power law acceleration for $c < \sqrt{3}$.
      Specifically, for $\tilde{b} \lesssim r_c^{-1}$, the universe
      starts to accelerate at early times and asymptotes to zero
      relatively quickly, while, for $\tilde{b} {\gg} r_c^{-1}$,
      it starts to accelerate only after a certain proper time, which
      may last for an infinitely long period before approaching a state of
      zero acceleration.

\FIGURE[ht]{ \epsfig{figure=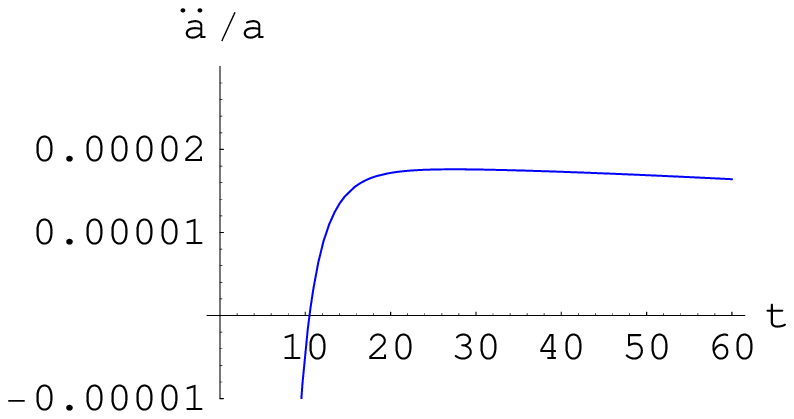, height=3.5cm, width=6.0cm}
        \hskip1.5cm
    \epsfig{figure=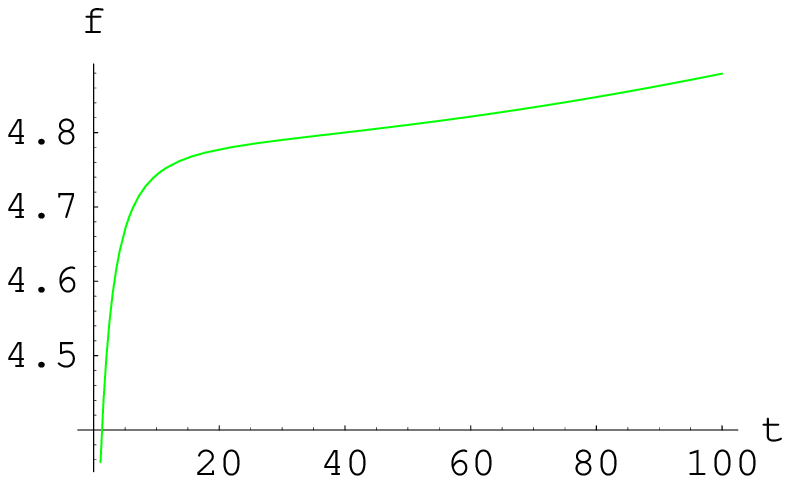, height=3.5cm, width=6.0cm}
    \caption{\sl (Left panel) acceleration vs
      $t$, and (right panel) the growth factor
      $f\equiv {f}(\varphi(t))$ vs proper time ($t$). The
      parameters are fixed at $r_c=1$, $\tilde{b}=1$,
      $k=k_1=-1$ and $c=3/\sqrt{7}$. The initial values 
      assigned at $t=0$ are $\varphi_0=0.5$, $\dot{\varphi}_0=0.6$ 
      and $a=1$.}
    \label{figure3}}

 \subsection{Compactification with a bulk cosmological term}

      From a
      pure gravity action with the $(4+m)$ dimensional cosmological
      term $\Lambda$, upon the dimensional reduction, we find that
      the 4d scalar potential is
      \begin{equation}
    V =-\,k_1
    {M_P^2}{{M}^2}\, {\rm e}^{- 2c\,\frac{\varphi}{M_P}}+
    \frac{1}{2}\,\tilde{\Lambda} M_P^2\,{\rm e}^{-\,\frac{2}{c}\,
      \frac{\varphi}{M_P}}\,,
    \label{M-poten1}
      \end{equation}
      where $\tilde{\Lambda}\equiv \Lambda (4/m(m-1))^{1/c^2}$ with
      $m\geq 2$. The field equations then have the following form:
      \begin{eqnarray}
    &&\ddot{\varphi}+3H\dot{\varphi}+ {2 c\, k_1}{{M}^2}\,
    {\rm e}^{-2c\,\varphi}
    -\frac{\tilde{\Lambda}}{c}\, {\rm e}^{-(2/c)\varphi}=0\,,
    \label{waveeqn2}\\
    && 3H^2-\dot{\varphi}^2+ {2k_1}{{M}^2} {\rm
      e}^{-2c\,\varphi}+\frac{3k}{a^2}-\tilde{\Lambda} {\rm
      e}^{-(2/c)\varphi}=0\,,\label{Friedman2}
      \end{eqnarray}
      in units $\kappa=1$. Here the value
      $c=1$ is transitional. The consistency
      equation~(\ref{identity}), which has no dependence
      on the flux parameter $b$, is the same.
      When $k=0$ and $k_1<0$ (again we take $k_1=-1$, for
      simplicity), we can find the exact solution for $c\approx 1$,
      which is given by
      \begin{eqnarray}
    &&\sqrt{6}\,\alpha^{\prime} = {M} \gamma \left(x+x^{-1}\right),
    ~~\quad
    \sqrt{2}\,\varphi^{\prime} ={M}\gamma \left(x-x^{-1}\right),
    \nonumber \\
    && x\equiv \frac{\sqrt{3}+1}{\sqrt{2}}\,
    \tanh \left({M}\gamma \tau\right), ~~\quad \gamma\equiv
    \sqrt{1+\frac{\tilde{\Lambda}}{2{M}^2}}
  \end{eqnarray}
    up to a shift of $\tau$ around $\tau=0$. Interestingly, this
    solution is accelerating for all $\tau$ satisfying
      \begin{equation}
    \tau  > \sqrt{\frac{2}{2M^2+\tilde{\Lambda}}} \cosh^{-1}
    \sqrt{\frac{3+2\sqrt{3}}{6}}.
      \end{equation}
      Acceleration is possible in the
      both cases $\Lambda>0$ and $\Lambda<0$, given that
      $\gamma>0$ holds. This translates to the
      condition that $\tilde{\Lambda}>-\,2{M}^2$. As in the case
      of form-field background, the accelerated
      expansion of a spatially flat universe can be eternal
      only if $c\leq 1$, otherwise the acceleration is transient.
      However, for the $k=-1$ cosmology, there can arise an eternally
      accelerating expansion even if $c \gtrsim 1$.

\section{Compact hyperbolic manifolds and mass gap}

     Here we would like to elaborate a little more on the
     issue of size of the internal space and its associated mass gap.
    In our model, it may look somewhat
    disappointing that under the time-evolution not only
     the size of the physical 3-space but also the scale
     factor of the internal space expands. We argue that this
     itself is not
     a real problem, but instead may be a desirable feature
     if one really wants to achieve a slowly varying
     4d scalar potential from the time-dependent string or M theory
     compactification on hyperbolic spaces.

In~\cite{GKL03a}, it was argued that the $m_{KK}\sim
10^{-60}$ (in 4d Planck units), a value closer to the current
Hubble scale $H_0\sim 10^{-33}~\mbox{eV}$, which is
phenomenologically unacceptable. If so, the hyperbolic
compactification in time-dependent background may be less
attractive. Here we revisit the argument and show that it should
be modified.

The mass gap discussed in~\cite{GKL03a} was based on the
expectation that the Kaluza-Klein masses are bounded below by
\begin{equation}\label{kkmass}
m_{KK} ={\cal O}(1)\, {\rm e} ^{-(m+2)\phi/2}\,r_c^{-1}.
\end{equation}
This estimate for a mass gap seemed to come from two more,
somewhat independent, assumptions: firstly, the mass gap on a CHM
is solely set by the scale $r_c^{-1}$, and secondly, in a
(conformal) frame appropriate to the 4d observer, the metric is
\begin{equation}\label{nonconf1}
d\hat{s}_{4+m}^2 = g_{\mu\nu}(x) dx^\mu dx^\nu
 + r_c^2\,{\rm e}^{(m+2)\phi} d\Sigma^2_{m, k_1}.
\end{equation}
These both points will require further scrutiny.

Let us recall that the metric solution that we have found 
(cf section 2) is
\begin{equation}\label{conformal1}
ds_{4+m}^2 = e^{-m\phi} g_{\mu\nu}(x) dx^\mu dx^\nu
 + r_c^2\,{\rm e}^{2\phi} d\Sigma^2_{m, k_1}.
\end{equation}
The difference between (\ref{nonconf1}) and (\ref{conformal1}) is
significant. In (\ref{conformal1}), the estimate for the size of
the internal manifold as $r_c\, {\rm e}^{(m+2)\phi/2}$ is not
four-dimensional, since here one is not going to a conformal frame
appropriate to the 4d observer but instead to a $(4+m)$ metric
spacetime conformally related to (\ref{nonconf1})~\footnote{One
might as well make a distinction between~(\ref{conformal1}) and a
metric ansatz $ds_{4+m}^2= {\rm e}^{-m u(y)}ds_4^2 +{\rm
e}^{2u(y)} \tilde{g}_{ab}dy^a dy^b$ that one often makes in a
warped string theory compactification, where the harmonic function
$u$ has no time-dependence. In~(\ref{conformal1}), the (conformal)
factor ${\rm e}^{-m\phi}$ multiplies not only the scale factor of
the physical $3$-space but also the time differential $dt$. In
turn, ${\rm e}^{-m\phi/2}\,a(t)$ is not the scale factor for the
proper time $t$, but it is, approximately, for the time $\tilde{t}
=e^{-m\phi/2}\,t$.}. In addition, there is no {\it a priori}
reason that $m_{KK}\sim \sqrt{V(\phi)}$ since the mass gap has
anything to do with the eigenmode associated with the first linear
eigenfunction on the (compact) internal manifold. In any case, we
are not going to proceed in this manner.

Instead, we make the following observation. To evaluate a mass
gap, one must write the metric applicable to a 4d observer. In
doing this, the (conformal) factor $e^{-m\phi(t)}$ is absorbed
while (dimensionally) reducing the term $\sqrt{-g_{4+m}}/G_{4+m}$
into $\sqrt{-g_4}/G_4$. The mass gap is then a product of the
eigenvalue associated with first radial eigenfunction on the CH
$m$-manifold times a suppression factor. To a 4d observer, this
factor is simply
$$ {f} = \left(\frac{m(m-1)}{4}\right)^{1/(m+2)}
    \exp\left(\frac{2}{\sqrt{m(m+2)}}\,
    \frac{\varphi-\varphi_0}{M_P}\right).$$
It is known that in compact hyperbolic manifolds the first
eigenvalue $\mu_1$ (associated with Laplace-Beltrami operator)
cannot be bounded by either volume (specifically,
$1/\mbox{Vol}(\Sigma)^{1/m}$) or the linear
dimension~\footnote{Also called ``diameter'' $d$, which is defined
to be the maximum geodesic distance between two arbitrary points
on $H^m/\Gamma$.} alone~\cite{Cheng75a} (see
also~\cite{Starkman00a} which contain many of the points about the
spectrum of the CHMs, including the fact that the gap is not given
by  the volume but by the linear diameter of the CHM, and that
this could make the gap much bigger)~\footnote{I wish to thank G.
Starkman for a nice illustration about this point.}.

In the zero-flux case, the geometric lower bounds on massive
Kaluza-Klein excitations may be approximated, in the
time-dependent background, by (see the appendix)
\begin{equation} m_{KK}\sim \frac{{\cal O}(\pi)}{r_c\times {f}}.
\end{equation}
The KK modes around $\varphi\sim \varphi_0$ can be extremely
heavy, $m_{KK}\gtrsim {\cal O}(\pi) r_c^{-1}\simeq {M}\sim
10^{15}\,\mbox{GeV}$ and their number can be low $\simeq
\frac{M_P^2}{{M}^2}\leq 10^{6}$.

  The question arises as to whether it is possible to
  use M-theory motivated potential for dark energy, namely,
  $V(\varphi) \sim 10^{-120}\,M_P^4$. First, one might note that
  the Hubble parameter we
  measure is purely a 4d quantity, and so even the approximation
  $H\sim \sqrt{V}$ (assuming that $k=0$ and
  $\dot{\varphi}^2<<V(\varphi)$)
  may be valid only if one expresses the
potential in terms of a canonically normalized 4d scalar
$\varphi$. It is suggestive to analyze the results in terms of 4d
cosmological parameters. For the M theory case, $m=7$, we have
\begin{equation}
V(\varphi)={{M}^2 M_P^2} {\rm
e}^{-\frac{6}{\sqrt{7}}\,\frac{\varphi-\varphi_0}{M_P}}
+\frac{M_P^2\,\tilde{b}^2}{2}\,e^{-2\sqrt{7}\,\frac{\varphi-\varphi_0}{M_P}}.
\end{equation}
For the tentative value $M\sim 10^{-3} M_P\sim
10^{15}\,\mbox{GeV}$, to get $V(\varphi)\sim 10^{-120} M_P^4$, we
require $\varphi-\varphi_0\sim 115.7$ (in 4d Planck
units)~\footnote{We realized that the scale $M$ (or $r_c^{-1}$) we
considered in the earlier version of this paper was too low.}.
This helps to estimate the growth factor ${f}$ by which the size
of the compact internal space may have increased. To a 4d
observer, the growth factor is ${f} \sim 6\times 10^{12}$ and
hence $(r_c\times {f})^{-1}\sim 166\,{GeV}$, a mass gap that is
physically interesting. It is plausible (but certainly not
established) that when $V(\varphi)$ dropped from its Planck scale
value $\sim 10^{66}\,(\mbox{GeV})^4$ to the current value
$V\approx V_0= 10^{-120}\,M_P^4$, the mass gap would decrease from
$\sim 10^{15}\,\mbox{GeV}$ to $ {\cal O}({10}^2)\,\mbox{GeV}$.

Let us consider another example, which is based on the exact
solution that we have presented in subsection 2.4. For
$c=3/\sqrt{7}$, we easily compute
         \begin{equation}
          {f} = \exp\left(\frac{2\,\varphi}{\sqrt{63}}\right)
       (10.5)^{1/9} \simeq  {1.29}\
            \left(\frac{3}{\sqrt{7}}\,{M}t \right)^{2/9}\,.
      \end{equation}
When $M=10^{-3} M_P$ and $t$ is as large as inverse of the present
Hubble scale $t_0\sim H_0^{-1} \sim (10^{-33}\,\mbox{eV})^{-1}$,
one finds $(r_c\times {f})^{-1}=
       R_c^{-1}\simeq 1.62\times 10^{11}\,\mbox{eV}\simeq
       162~\mbox{GeV}$. This mass gap must clearly
       be considered tentative until the size of compactification
      around $\varphi\sim \varphi_0$, where
      ${f}\sim {\cal O}(1)$,
      or the curvature radius of internal space
      $r_c$ ($\equiv M^{-1}$) is known.

       Of course, one could argue that even if
       $m_{KK}\sim \mbox{TeV}$ in these cases, one finds
       $m_{KK}<< \mbox{TeV}$ when $\varphi-\varphi_0>> 115$
       and $t >> t_0 \equiv {10}^{33} (\mbox{eV})^{-1}$. The issue is
      whether we are willing to accept the value
 $V << V_0 \approx 10^{-120} M_P^4$.  For example, in order
 to witness $m_{KK} \sim 10\,\mbox{GeV}$, we require
${f}\sim 10^{14}$ and hence $t\gtrsim 2.8\times
{10}^{38}\,(\mbox{eV})^{-1}$, a time scale which is too large,
 like a few trillion years, and we may practically never approach
 this limit.

  \section{Discussion}

      In this paper, we have used
      the M-theory motivated scalar potentials to study the FRW type four
      dimensional flat (and open) universe cosmologies and shown that
      a scalar potential of the form
      $V \sim  \Lambda_1 \exp (-2c\kappa \varphi)+ \Lambda_2
      \exp (-(6/c)\kappa\varphi)$
      arising from the hyperbolic-flux
      compactification might be of interest for getting transiently or
      eternally accelerating expansion of the universe. For a spatially
      flat universe the accelerated
      expansion will continue forever only if the coupling
      $c\leq 1$, regardless of the background flux.
      In light of the result that for all known classical
      compactifications of supergravity theories on some non-trivial
      curved internal manifolds or toroidal spaces with trapped fluxes, only
      $c\gtrsim 1$ arises in practice, we are led to explore other
      alternatives for cosmic acceleration. We find that an eternally
      accelerating expansion is possible with $c>1$ if the
      spatial curvature of the universe is negative. For the
      case of hyperbolic compactification, this claim
      was made before in ref.~\cite{CHNOW03b} based on a result
      from linearized approximations. Here we have also generalized
      those results by solving
      the classical field equations with non-zero fluxes.

    It is known, at least, for $m=2$ and $m=3$, that the most compact
    manifolds are hyperbolic. In the context of hyperbolic
    compactification, we point out that the existence of
    a mass gap in the ${\cal O}(10^2)\, \mbox{GeV}$
    is not impossible, since the curvature radius of the internal
    space can be extremely small, like $r_c \sim {\cal O}(10) M_P^{-1}$.

    One direction of further investigation is to analyse the
    field equations as a dynamical system, in the manner of Halliwell,
    using a phase-plane method and see how the solutions behave
    in phase space asymptotically in the combined case of
    the hyperbolic-flux compactification. A step in this
    direction, with the settings $r_c=1$ and ${b}=0$, is
    recently taken in~\cite{Pedro} (see also the last
    section of ref.~\cite{Miguel03a}), which generalize the
    earlier phase-plane analysis in ref.~\cite{Halliwell87a}.

    Another direction is to extend our construction by taking
    some of the extra dimensions to be flat,
    like $T^4\times H^3/{\Gamma|_{free}}$, instead of the
    $H^7/\Gamma$ geometry, since the energy spectrum on
    a CH-$3$ manifold is better known as compared a
    CHM in four and higher-dimensions. In addition, every hyperbolic
    manifold $\Sigma_{m}=\H^m/\Gamma$ of finite volume, with
    $m\geq 3$, may be decomposed disjointedly into a relatively
    compact $\Sigma_0$ and many finite cusps.
    One may also include the brane (or brane-instanton)
    effects by allowing a delocalized flat
    transverse space. In this case, even though the
    flat part will have zero contribution to a scalar potential,
    the coupling constant $c$ will change. We hope to return to this
    and related problems elsewhere.

\section*{Acknowledgments}

It is a pleasure to thank R Emparan, Pei-Ming Ho, K T Inoue, T 
Mohaupt, N Ohta, S Randjbar-Daemi, B Ratra, G Starkman, P K 
Townsend and J Weeks for discussions and helpful comments. The
author also thanks C-M Chen and J Wang for earlier discussions,
and M Costa, L Jarv and F Saueressig for helpful
correspondences. He also acknowledges the warm hospitality of the
Abdus Salam ICTP during the period when part of this work was
being completed. This research was supported in part by the
National Science Council and the CosPA project of the Ministry of
Education, Taiwan.

\section*{Appendix: Some useful results on CHMs}
\renewcommand{\theequation}{A.\arabic{equation}}
\setcounter{equation}{0}

The eigenvalue equation
        $\left(\Delta_{LB} +\mu_n^2\right) {\cal U}_{\mu}=0$
       of the Laplace-Beltrami operator $\Delta_{LB}$ on a Riemannian
    manifold possessing both local geometry and global topology plays
    a significant role to set the scale of the metric
    perturbation. There
    may exist many bounds for the $n$th eigenvalue ${\mu_n}~(\equiv
   \sqrt{E_n}$) in terms of the diameter $d$ and the
    Ricci curvature $\tilde{R}$ of the manifold $\Sigma$ itself. For our
    purpose, it is crucial to know the behavior of
    low-lying eigenmodes.

As for m-dimensional compact hyperbolic manifolds, there are a
number of estimates of $\mu_1 (=\sqrt{E_1})$ in mathematical
literature, see, e.g.,~\cite{Cheng75a}. If the Ricci curvature
$\tilde{R}(\Sigma)$ is bounded below by $-\,(m-1)/l^2$, then
$\mu_1 (=\sqrt{E_1})$ satisfies
\begin{equation}
\mu_1^2 \geq \mbox{max}\left[\frac{\pi^2}{2d^2}-\frac{1}{4 l^2},
\sqrt{\frac{\pi^4}{d^4} +\frac{1}{16 l^4}}-\frac{3}{4l^2},
\frac{\pi^2}{d^2} \exp \left(-c_m \frac{d}{2 l} \right)\right],
\end{equation}
where $c_m=\sqrt{2}$ if $m=2$ and $c_m=\sqrt{m-1}$ if $m\geq 3$.
$\mu_1$ determines the maximum fluctuation scale of the
perturbation and hence the geometric lower bound on KK masses. If
$d$ is larger than $\sqrt{2}\pi l$, then $E_1>0$ but it does not
say anything about the existence of modes $\mu<1/r_c$. At any
rate, since $\mbox{Vol}(\Sigma)\sim r_c^m e^\alpha$, it is clear
that the first eigenvalue $\mu_1$ cannot be bounded by either
volume or the diameter alone.

It is also known that, for $m\geq 2$, there exist upper bounds to
the eigenvalues $\mu_n$ (counted with multiplicity,
$0=\mu_0(\Sigma)<\mu_1(\Sigma) \leqq \mu_2(\Sigma) \leqq \cdots
$)~\cite{Cheng75a} (there is an incorrect sign in~\cite{Tabbash01a})
\begin{eqnarray}
&&  \mu_n^2 \leq \left\{\begin{array}{l} \frac{4\pi^2 n^2}{d^2}
\left(1+2^{(m-2)/2}\right)^2+\frac{(m-1)^2}{4\,l^2}, \quad
(m=\mbox{even}),
\\
\frac{4(1+\pi^2) n^2}{d^2} \left(1+2^{m-3}\right)^2
+\frac{(m-1)^2}{l^2}, \quad  (m=\mbox{odd}).
\end{array} \right.\
    \end{eqnarray}
$\mu_1$ may be arbitrarily close to zero if $m=2$. This is
understandable because one can deform a CH$-2$ manifold
continuously so that $d\to \infty$. For CHM with $m\geq 3$, this
is prohibited and by Mostow-Prasad rigidity theorem there are no
massless shape moduli~\cite{Starkman00a}. For the $m=3$ case, a
more useful empirical relation is the Weyl formula:
    \begin{equation}
    {\mu}_1=
    \sqrt{\frac{1}{r_c^2}+\left(\frac{9\pi^2}{\mbox{Vol}(\Sigma)}\right)^{2/3}}.
    \end{equation}
The mass gap can much larger than $(\mbox{Vol}(\Sigma))^{-1/m}$
when the `complexity' of the manifold (i.e., the $m$-dimensional
generalization of the genus) is large. The above relation was
found in close agreement with a numerical result for ``smallest''
$263$ CH-$3$ manifolds~\cite{Inoue01a}.

In the limit $d>> r_c/2$, where the manifold converges to the
original cusped manifold and $\mu_1(\mbox{cusp})\sim 1/r_c$ , one
has
$$
\mu_1 = \sqrt{(\mu_1(\mbox{cusp}))^2+\frac{4\pi^2}{\beta^2 d^2}},
$$
and $ \mbox{Vol}(\Sigma)\simeq \mbox{Vol}(\Sigma_{\mbox{cusp}})
[1- \exp(-2(d-d_0))]$ where $d\geqq d_0=0.25\,r_c$. Here $d_0$ is
the diameter of the complementary part to that of a `thin' part.
One may cut the neighborhood of a cusp (or `thin' part), so that
$d_0$ becomes the usual diameter. This relation holds for some
manifolds with one `thin' part (see~\cite{Inoue01a} and
references therein). The numerical results in~\cite{Inoue01a} also
suggest that for `smallest' CH-$3$ manifolds
$\langle\beta\rangle=1.7$, although for each manifold the value of
$\beta$ may be different, e.g., $\beta=1.3$ is the value for a
specific manifold.

Another method to compute the eigenvalues (in terms of periodic
orbits) is the Selberg trace formula, e.g., for CH-$2$ manifold,
$\mu_1> 0.47/r_c$~(see, e.g.~\cite{Starkman00a} and references
therein). This method has not been used to any CHMs with $m\geq
3$. The Weyl asymptotic formula that generally holds for modes
with $\mu_1>> 1/r_c$ gives the number density of KK states:
$$ {\cal N}(\mu)\sim \frac{1}{(2\pi)^m} \,\Omega_m
\mbox{Vol}(\Sigma_m) \mu^m, $$ where
$\Omega_m=2\pi^{m/2}/(m\,\Gamma(m/2))$ is the volume of the unit
disk in a Euclidean $m$-space. The formula written 
in~\cite{Starkman00a} (cf equation~(8)) may give a good
approximation for CHMs that extend in one direction, i.e., $d>>
r_c/2$.

There can be only a lower bound to the volume of compact
hyperbolic space, like $V_{CH3}> 0.1667\, r_c^3$, and the known
example with smallest volume is Weeks manifold with $V>
0.94\,r_c^3$. For some specific $m=2$ or $m=3$ CH manifolds, it
often happens that $e^{\alpha}\sim {\cal O}(1)$. In general,
however, the quantity $e^{\alpha}$ is arbitrary, which is
essentially a counting of the number of handles on the CH
manifold~\footnote{I wish to thank K.T. Inoue and G. Starkman for
detailed discussions of this and related issues.}. If one allows
the internal space to be compact hyperbolic orbifolds then the
volume can be much smaller, see, e.g.,\cite{Inoue}.


\end{document}